\documentclass[aps,prl,twocolumn,superscriptaddress,reprint]{revtex4-1}

\usepackage{amsmath}
\usepackage{amsfonts}
\usepackage{graphicx}
\usepackage{tikz-feynman}
\usepackage{cancel}

\usepackage[colorinlistoftodos,prependcaption,textsize=small]{todonotes}

\makeatletter

\newcommand{\mune}{\mu_{\text{NE}}}
\newcommand{\mue}{\mu_{\text{E}}}

\newcommand{\mt}{\mathfrak{t}}

\makeatother

\begin{document}
	\title{Stealth entropy production in active field theories near Ising critical points}
	
	\author{Fernando Caballero} 
	\affiliation{DAMTP, Centre for Mathematical Sciences, University of Cambridge, Wilberforce Road, Cambridge CB3 0WA, UK}
	
	\author{Michael E. Cates} 
	\affiliation{DAMTP, Centre for Mathematical Sciences, University of Cambridge, Wilberforce Road, Cambridge CB3 0WA, UK}
	
	\date{\today}
	
	\begin{abstract}
		We address the steady-state entropy production rate (EPR) of active scalar $\phi^4$ theories, which lack time-reversal symmetry, close to a phase-separation critical point. We consider both nonconserved (Model A) and conserved (Model B) dynamics at Gaussian level, and also address the former at leading order in $\epsilon = 4-d$. In each case, activity is irrelevant in the RG sense: the active model lies in the same (dynamic Ising) universality class as its time-reversible counterpart. Hence one might expect that activity brings no new critical behavior. Here we show instead that, on approach to criticality in these models, the singular part of the EPR per (diverging) spacetime correlation volume either remains finite or itself diverges. A nontrivial critical scaling for entropy production thus ranks among universal dynamic Ising-class properties.	
	\end{abstract}
	
	\maketitle

	In active materials the individual particles are out of equilibrium locally, constantly consuming energy that is transformed into motion via self-propulsion \cite{ramaswamy2010mechanics, ramaswamy2017active, Marchetti2013RMP}. Activity gives rise to new physics that would not be observed in passive equilibrium systems, like flocking \cite{Vicsek1995PRL,toner1998flocks,Toner1995,Solon:2015:PRL}, and motility-induced phase separation (MIPS) \cite{Cates2015,Tailleur:08,Speck:13,stenhammar2014phase,HenkesABP,Fily:12,PhysRevLett.108.248101}. In MIPS, separation into dense and dilute fluid phases (denoted liquid and vapor) arises without attractive interactions. This is because collisions between particles heading in opposite directions have a long lifetime -- the orientational decorrelation time -- that mimics an attractive pair interaction without detailed balance \cite{Cates2015,HenkesABP,Fily:12}. 
	
	Coarse-graining of the microscopic dynamics of MIPS gives continuum models \cite{Tailleur:08,Cates2015,stenhammar2014phase,PhysRevLett.111.145702,PhysRevLett.112.218304} which can be distilled into canonical form as dynamical $\phi^4$ stochastic field theories \cite{wittkowski2014scalar,Nardini2017,tjhung2018reverse}. These canonical theories are  amenable to renormalization group (RG) study \cite{caballero2018strong,caballero2018}, not least because of the absence of multiplicative noise. They capture much of the known phenomenology of MIPS, at least in the absence of hydrodynamic interactions (for which see \cite{PhysRevLett.115.188302,PhysRevLett.123.148005}), including activity-induced violations of the common-tangent construction for phase equilibria \cite{wittkowski2014scalar,PhysRevE.97.020602}, and microphase separation into phases of finite clusters surrounded by vapor and/or finite vapor bubbles within a dense liquid \cite{tjhung2018reverse}.
	
	Crucially, and distinctively, these active $\phi^4$ theories contain terms that break time reversal symmetry (TRS). In the simplest cases this is via a contribution to the local chemical potential whose form prohibits the construction of a global free energy functional \cite{wittkowski2014scalar}. Because the order parameter $\phi$ is dynamically conserved (but see \cite{Cates11715,PhysRevLett.119.188003,PhysRevE.92.012317,Weber_2019} for exceptions) the models so far studied are mainly extensions of (passive) Model B; they are known as Active Model B (AMB) and Active Model B+ \cite{wittkowski2014scalar,Nardini2017,tjhung2018reverse}. (Recall that passive Model B describes the conserved, diffusive dynamics, with TRS, of a scalar field governed by a free energy with a symmetric $\phi^4$ local term and a square gradient correction \cite{ChaikinPaulM2000Pocm,hohenberg1977,goldenfeld2018lectures}.) However, to help understand the effects of activity in fundamental terms, we introduce and study below a simpler, non-conserved variant (Active Model A, or AMA), again with a TRS-breaking chemical potential contribution. 
	
	In this Letter we address the effects of activity on universal behavior close to a liquid-vapor critical point. Although activity can create a richer phase diagram (featuring strong coupling regimes \cite{caballero2018} and/or microphase separation \cite{tjhung2018reverse}), our previous work on conserved systems close to and above dimension $d=4$ establishes that the leading order active terms responsible for TRS breaking in our active $\phi^4$ field theories are {\em irrelevant}, in the RG sense, near the usual liquid-vapor critical point. It follows by definition that, despite activity, such models then fall within the conserved (Model B) or nonconserved (Model A) dynamic Ising class \cite{caballero2018}.
	
	We now ask the following question: {\em Given that activity is irrelevant, does this mean that the critical behavior of an active model is, effectively, passive?} Our answer, surprisingly, is no. This question is significant, not just for the liquid-vapor critical point, but for other cases where active models appear to lie within a passive universality class (see, {\em e.g.}, \cite{PhysRevLett.112.218304,LeeNat,Marchetti2013RMP}). To arrive at our negative answer, we compute the RG flow of the steady-state entropy production rate (EPR) of active $\phi^4$ models. We do this near the Gaussian fixed point (GFP, which exists in all dimensions and is stable for $d>4$) for both AMA and AMB, and also near the fully nonlinear Wilson-Fisher fixed point (WFFP) to order $\epsilon=4-d$ in the simplest (nonconserved) case of AMA only. We first identify a natural scaling for EPR in active systems close to criticality, and show that near the GFP this scaling still holds, despite the formal irrelevance of activity. Remarkably, in $d=4-\epsilon$ we find that the critical EPR close to the WFFP is much {\em larger} in magnitude than the natural scaling would suggest. Notice that any equilibrium system has a vanishing EPR, so that our results only give information about a system whose activity is strictly nonzero, while remaining irrelevant in the RG sense.
	
	Before giving our formal calculations of the steady-state EPR per unit volume, which we denote $\sigma$, we first identify its `natural' scaling, as just referred to. Note first that we are interested only in the singular part of $\sigma$ (connected with critical phenomena) just as, for static properties in equilibrium models, we are interested only in the singular part of the free energy density $f = - \ln Z/V$. Here $Z=\int e^{-{\mathcal F}[\phi]}\mathcal{D}[\phi]$, with $\mathcal{F}$ the free energy functional. This allows us to discard all terms independent of the system's correlation length,  $\xi\sim \xi_0\mt^{-\nu}$, where $\mt$ is the dimensionless control parameter (in equilibrium, the `reduced temperature') for the transition. In particular, all terms that depend {\em solely} on the short cutoff, $\xi_0$ (not on $\mt$) can be ignored, even if they diverge as $\xi_0\to 0$. 
	
	For a static equilibrium system near criticality, $f \sim \xi^{-d}$ \cite{ChaikinPaulM2000Pocm,AmitBook}. This result lies behind the hyperscaling relation for the specific heat exponent: $c_V = -d^2f/d\mt^2\sim\mt^{-\alpha}$ with $-\alpha=\nu d- 2$.  Note that it applies not only for the WFFP in Ising-class models, but also for the unstable GFP found in $d<4$ by tuning to zero the $\phi^4$ term in ${\mathcal F}$. (This tuning accesses a tricritical point in $3<d<4$ \cite{ChaikinPaulM2000Pocm}.) In $d>4$ the GFP is stable, and hyperscaling still holds for the fluctuation part of $f$, whose heat-capacity contribution is, however, now subdominant to the mean-field discontinuity. The robust hyperscaling result, $f\sim\xi^{-d}$, can be simply interpreted in terms of `block-spins' found by successive coarse-graining and rescaling through the Kadanoff transformation: once the coarse graining length reaches $\xi$, the block-spins become independent. Thus, for the purposes of $f$, there is one block-spin degree of freedom per correlation volume. 
	
	A simple extension of this reasoning argues that each block-spin flips once per correlation time $\tau\sim\tau_{0}(\xi/\xi_0)^{z}$, where $z$ is the usual dynamic exponent, and $\tau_0$ a microscopic time. So long as these block-flip events break detailed balance, we expect  an order-one entropy production per event. (We use informatic units: $k_B = 1$.) If so, the entropy production scales as $\sigma\sim\xi^{-(d+z)}\sim\mt^{\nu(d+z)}$. Just as with $f$, all terms regular in $\mt$ are discarded,  including any that diverge as $\xi_0\to 0$. This is because we are not interested here in the total calorific heat production, which is dominated by microscopic irreversibility near the short cutoff, but in the part of the EPR that quantifies irreversibility at large scales near criticality \cite{Nardini2017}.
	
	In this context, the fact that $\sigma$ vanishes (by the above argument) as $\mt\to 0$ does {\em not} make it physically insignificant. Indeed, $f\sim\xi^{-d}$ also vanishes, yet it is clearly significant, creating, {\em inter alia}, the $c_V$ divergence. We argue instead that an effectively reversible critical dynamics can emerge only if $\sigma$ vanishes {\em faster} than the natural scaling, $\mt^{\nu(d+z)}$. That is, if we define an exponent $\theta_\sigma$ via $\psi \equiv\sigma\xi^{d+z}\sim \mt^{-\theta_\sigma}$, where $\psi$ is now the steady-state EPR per spacetime correlation volume, the critical dynamics remain effectively active unless $\theta_\sigma$ is strictly negative, so that $\psi$ vanishes at the critical point.
	
	We find below that $\theta_\sigma$ is not negative but zero at the Gaussian fixed point for Active Models A and B.  Moreover, we show it is generically positive to order $\epsilon = 4-d$ for the Wilson-Fisher fixed point of Active Model A. Our results establish that irreversibility can remain important in the critical regime for active models, even when these share a dynamic universality class with reversible models. We term this `stealth' entropy production, and discuss it further at the end of the paper. 
	
	The models of interest each obey a Langevin equation for the field $\phi$ such that the probability of a path $\lbrace\phi(r,t)\rbrace_{t\in(0,\mathcal{T})}$  is defined by a distribution $\mathcal P[\phi]=e^{-{\mathcal A}_+[\phi]}$ with ${\mathcal A}_+$ a dynamical action detailed below. The global steady-state EPR can be written using results from stochastic thermodynamics (see \cite{Seifert2012}) as \cite{Nardini2017}:
	\begin{equation}\label{eq:entDef}
	\mathcal{S}=\lim_{\mathcal{T}\to\infty}\frac{1}{\mathcal{T}}\left\langle\ln\frac{\mathcal P[\phi]}{\mathcal P^R[\phi]}\right\rangle. 
	\end{equation}
	Here $\mathcal{P}^R[\phi] = e^{-{\mathcal A}_-[\phi]}$ is the probability of the time-reversed path under the forward dynamics, and the average is over noise realizations in the steady state. Note that $\cal S$ can be expressed in terms of the mismatch between response and correlator observables via a Harada-Sasa type integral \cite{Nardini2017,PhysRevE.73.026131}.
	
	Considering first AMA, the dynamical action takes the form \cite{Tauber2014}:
	\begin{equation}\label{eq:actionDef}
	\mathcal{A}_\pm[\phi] =\frac{1}{4D}\int d^dr\,dt\, (\dot\phi\pm\mu)(\dot\phi\pm\mu)
	\end{equation}
	where $D$ sets the variance of a Gaussian white noise obeying $\langle\eta(r,t)\eta(r',t')\rangle = 2D\delta^{(d)}(r-r')\delta(t-t')$ in the corresponding Langevin equation for $\phi$. This reads
	\begin{equation}\label{eq:Langevin}
	\dot\phi = -\mu +\eta
	\end{equation}
	in which $\mu = \mue+\mune$, where $\mue =\delta \mathcal{F}/\delta\phi$ is a chemical potential deriving from the usual $\phi^4$ free energy
	\begin{equation}\label{eq:F}
	\mathcal F = \int d^dr \,\frac{a}{2}\phi^2+\frac{u}{4}\phi^4+\frac{\kappa+2\kappa_1\phi}{2}(\nabla\phi)^2.
	\end{equation}
	Here the irrelevant (passive) $\kappa_1$ term must be included for RG purposes \cite{caballero2018}, because it interacts with active terms of the same order arising in $\mune$, as we see below. 
	
	By definition, $\mune$ is {\em not} of the form $\delta \mathcal{F}/\delta \phi$ for any $\mathcal{F}$, thereby breaking TRS. To leading order in $(\nabla,\phi)$ it can be written \cite{caballero2018}
	\begin{equation}\label{eq:mune}
	\mune = (\lambda+\kappa_1)(\nabla\phi)^2.
	\end{equation}
	This notation ensures $\mu=a\phi+u\phi^3-(\kappa+2\kappa_1\phi)\nabla^2\phi +\lambda(\nabla\phi)^2$ so that $\kappa_1$ and $\lambda$ multiply distinct nonlinearities. Note that $\kappa_1=0,\lambda\neq0$ in the specific active field theories of  \cite{wittkowski2014scalar,Nardini2017,tjhung2018reverse}, while equilibrium models have $\lambda+\kappa_1=0$.
	
	Substituting ${\mathcal A}_\pm$ into \eqref{eq:entDef}, and using the large $\mathcal{T}$ limit to discard terms in $\int d^dr\,dt\,\dot\phi\mue=\int dt\,\dot{\mathcal F}$, we find
	\begin{equation}\label{eq:entGen}
	\begin{aligned}
	\sigma &= \frac{\mathcal{S}}{V}=\frac{-1}{DV}\int d^dr \langle\mune\dot\phi\rangle\\ &= -\langle\mune\dot\phi\rangle/D = \langle\mune(\mu-\eta)\rangle/D
	\end{aligned}
	\end{equation}
	where we assume translational invariance so the spatial integral cancels the factor $1/V$.
	It is easily shown that the corresponding results for AMB are the same as (\ref{eq:actionDef}--\ref{eq:entGen}) after substituting $(\mu,\eta)\to(-\nabla^2\mu,-\nabla\cdot\vec\eta)$, with $\vec\eta$ a $d$-component Gaussian white noise of variance $D$, in (\ref{eq:actionDef},\ref{eq:Langevin}) and in the final member of \eqref{eq:entGen} \cite{Nardini2017,caballero2018}.

	Because the coupling constants $\lambda$ and $\kappa_1$ are RG-irrelevant, they vanish at the fixed point. The critical behavior of $\sigma$ is then computed in two stages, finding first the flow of the coupling constants themselves, and second that of the terms they multiply within the expression for $\sigma$ in \eqref{eq:entGen}. These terms are composite operators, whose scalings can be calculated within the critical-point theory directly \cite{AmitBook}, where $\phi\to-\phi$ symmetry is recovered. Averages containing odd powers of $\phi$ thereby vanish, along with the term multiplying the noise, since it represents the average of an odd number of Gaussian noises when expanded close to the WFFP.
	
	It remains to find the RG flow of the even terms, which for AMA are
	\begin{equation}\label{eq:sigmamune}
	\sigma = D^{-1}\langle\mune(\lambda(\nabla\phi)^2-2\kappa_1\phi\nabla^2\phi)\rangle_{\text{cr}}
	\end{equation}
	where the average $\langle\cdot\rangle_{\text{cr}}$ is over the stationary measure of the passive $\phi^4$ theory at criticality. We now split this entropy production into three terms, $\sigma = \sigma_1+\sigma_2+\sigma_3$, such that the coefficients of each term factorize into $\lambda$ and $\lambda+\kappa_1$ (which have pure scalings in the RG flow as exemplified in  \eqref{eq:KTA} below):
	\begin{equation}\label{eq:sigmaA}
	\begin{aligned}
	\sigma_1 &= D^{-1}\lambda(\lambda+\kappa_1)\langle(\nabla\phi)^4\rangle_{\text{cr}},\\
	\sigma_2 &= -2D^{-1}(\lambda+\kappa_1)^2\langle(\nabla\phi)^2\phi\nabla^2\phi)\rangle_{\text{cr}},\\\sigma_3 &= 2D^{-1}\lambda(\lambda+\kappa_1)\langle(\nabla\phi)^2\phi\nabla^2\phi)\rangle_{\text{cr}}.
	\end{aligned}
	\end{equation}
	Notice that, written this way (effectively by adding $0=2(\lambda-\lambda)\phi\nabla^2\phi$ to the parenthesis in \eqref{eq:sigmamune}) all three terms vanish separately in equilibrium where $\lambda+\kappa_1=0$. Note also that these terms can be calculated perturbatively close to the equilibrium fixed points, where the active parameters are irrelevant and thus asymptotically small.
	
	The composite operators in $\sigma$ have dimension $\nabla^{4}\phi^4$. (We discard their non-critical UV divergences via dimensional regularization \cite{AmitBook}.) These are the only terms that survive when, close to either the GFP or the WFFP, we recover $\phi\rightarrow-\phi$ symmetry via vanishing $\lambda,\kappa_1$. This limit also allows us to neglect the noise term in the entropy production, since the noise is just the zeroth-order field and so gives odd terms when multiplied by $\mune$ \cite{thisSM}.
	
	Recall that the Kadanoff transformation integrates over a momentum shell $(\Lambda,\Lambda/b)$, then rescales momenta $q\to qb$, frequencies $\omega \to b^z\omega$ and fields $\phi\to b^{-\chi} \phi$. The correlation length flows as $\xi \to \xi/b$. By holding $D$ and $\kappa$ constant under this RG flow, the Gaussian and Wilson-Fisher  fixed points ($\xi\to\xi = \infty$) can be accessed, and $z$ and $\chi$ identified for each; see \cite{Tauber2014} for the procedure. 
	
	The flows are simple near the GFP, where the quartic coupling $u$ in \eqref{eq:F} either flows to zero (for $d>4$) or is held there by fiat, so as to access a tricritical point in $3<d<4$ \cite{ChaikinPaulM2000Pocm,AmitBook}. The terms in $\sigma$  (Eq.~\ref{eq:sigmaA}) are each of the schematic form $g_ig_jD^{-1}\nabla^{4}\phi^4$ with $g_i = (\lambda,\lambda+\kappa_1)$. The effects of the Kadanoff transformation are easily found:
	\begin{equation} \label{eq:KT}
	\begin{aligned}
	g_i&\to b^{z+\chi-2}g_i,\\
	D&\to b^{z-d-2\chi} D = b^0D\\
	\nabla^{4}\phi^4&\to b^{4-4\chi}\nabla^{4}\phi^4
	\end{aligned}
	\end{equation}
	where the constancy of $D$ fixes $z = d+2\chi$, while that of $\kappa$ fixes $2\chi = 2-d$. (Inserting these equalities into the exponent for the $g_i$ confirms their irrelevance in $d>2$.) 
	Exactly analogous results hold for AMB except that (i) the schematic form is now $g_ig_jD^{-1}\nabla^{6}\phi^4$; (ii) an extra factor $b^{-2}$ appears in the $g_i,D$ scalings; (iii)
	the composite operators scale as $\nabla^{6}\phi^4\to b^{6-4\chi}\nabla^{6}\phi^4$; and (iv) $z = d+2+2\chi$.
	
	Combining the above scalings, we obtain for both the conserved and non-conserved dynamics
	\begin{equation}\label{eq:sigmarescaling}
	\sigma \to b^{d+z}\sigma
	\end{equation}
	which, since $\xi\to\xi/b$, gives $\sigma\sim\xi^{-(d+z)}\sim\mt^{\nu(d+z)}$. This is precisely the natural scaling promised earlier, corresponding to $\psi \equiv \sigma\sim\xi^{d+z}	\sim t^{-\theta_\sigma}$ with $\theta_\sigma = 0$. Activity is therefore significant near the GFPs of both AMA and AMB, although the same universality classes include passive members (such as passive Models A and B) for which $\sigma$ vanishes identically. 
	
	Although its derivation involves static averages $\langle\cdot\rangle_{\text cr}$ taken within a linear theory (the Gaussian model), the final result \eqref{eq:sigmarescaling} describes models whose active couplings are intrinsically nonlinear. This explains why the RG approach is necessary, and its outcome nontrivial. 
	
	In the neighborhood of the Wilson-Fisher fixed point, the calculation is more involved; we pursue it for AMA only. The principle is the same, which is to find the scalings of the coupling constants $g_i$, and also of the composite operators that they multiply in \eqref{eq:sigmaA}. The exponent for each factor now has corrections of order $\epsilon = 4-d$ which we calculate as usual in a perturbative, one-loop, RG \cite{Tauber2014,AmitBook}. For the couplings we find (see \cite{thisSM} for details)
	\begin{equation} \label{eq:KTA}
	\begin{aligned}
	\lambda&\to b^{z+\chi-2-\epsilon/3}\lambda,\\
	\lambda+\kappa_1&\to b^{z+\chi-2-\epsilon/4}(\lambda+\kappa_1).
	\end{aligned}
	\end{equation}
	Note that we have defined couplings as $g_i = (\lambda,\lambda+\kappa_1)$ rather than, say, $(\lambda,\kappa_1)$ because, as shown further in \cite{thisSM}, the chosen pair do not mix in the RG flow. (This is also why we decomposed the entropy production in \eqref{eq:sigmaA} above into terms with pure $g_ig_j$ coefficients.) The same method recovers the well known one-loop RG results for Model A, namely $\nu = 1/2+\epsilon/12+\mathcal{O}(\epsilon^2)$, $\chi=(2-d)/2+\mathcal{O}(\epsilon^2)$ and $z = 2 + \mathcal{O}(\epsilon^2)$ \cite{Tauber2014}.
	
	Calculating anomalous dimensions for the composite operators in \eqref{eq:sigmaA} is  more burdensome. The main complication is that operators of the same dimension can mix. The bare  dimension of a local operator with $m$ fields and $n$ gradients is, from dimensional analysis, $m(d-2)/2+n$. For operators of dimension $6$ at $d\to d_c = 4$ in $\phi^4$ theory, the calculation is done in \cite{AmitBook}. For those of dimension $8$ in a system of upper critical dimension $6$, see \cite{PhysRevB.15.4657}.  
	The situation might look simpler in $d=2$ where all composite operators decompose into a small number of primary fields \cite{CardyBook}, but in $d=2$ our active terms are not irrelevant, so we do not know the measure $\langle\cdot\rangle_{\text cr}$ in \eqref{eq:sigmaA}.
	
	We can make progress in $d=4-\epsilon$, where $\sigma$ in \eqref{eq:sigmaA} contains two operators of dimension 8 with 4 fields; there are 7 of these operators in total \cite{thesis,thisSM}, and all must be treated together. (To order $\epsilon$, or one loop,  the only operators that can mix with those in $\sigma$ must likewise be quartic in $\phi$.)
	The calculation starts by adding the full set of operators $A_i$ with source fields $a_i$ to the action: $\mathcal{A}\to\mathcal{A}+\int  a_iA_i$. Averages involving the $A_i$ can then be calculated through functional derivatives with respect to the $a_i$ \cite{AmitBook}.
	
	This computation \cite{thesis,thisSM} defines a matrix $M_{ij}$ that enters the differential RG flow equations for the sources as
	\begin{equation}
	\frac{da_i}{db} = \tilde da_i+M_{ij}a_j,
	\end{equation}
	where, after adding the insertions into $\mathcal A$, $\tilde d = -4+z$ is the naive scaling dimension of all the new parameters $a_i$, such that the added terms match the dimensions of everything else in the action.
	Here $M_{ij}$ (detailed in \cite{thisSM}) is proportional to $u$ and hence to $\epsilon$, once it is evaluated at the fixed point, where $u = u^*= \kappa^2\epsilon\Lambda^4/(9D\Omega_d) + \mathcal{O}(\epsilon^2)$, where $\Lambda$ is the cutoff and $\Omega_d$ a geometric factor (see \cite{thisSM}). $M_{ij}$'s left eigenvectors ${v}_i^{\alpha} = {\bf v}^\alpha$, with $\alpha = 1...7$, define a set of independent scaling exponents via the corresponding eigenvalues (anomalous dimensions) $\delta^\alpha$ as
	\begin{equation}
	\frac{d{\bf v}^\alpha}{db} = (\tilde d+\delta^\alpha){\bf v}^\alpha.
	\end{equation}
	
	As outlined in \cite{thisSM}, the eigenvalues are found as
	\begin{equation}\label{eq:eigenvalues}
	\delta^\alpha=-\epsilon\left(\frac{10}{9},1,1,\frac{13}{18},\frac{1}{2},\frac{1}{2},\frac{1}{3}\right).
	\end{equation}
	All are negative, so that the most relevant eigen-operators are those whose eigenvalues are of least magnitude. The leading scaling of any $A_i$ is that of the most relevant eigen-operator onto which it has nonzero projection. For $\sigma_{1}$ and $\sigma_{2,3}$ in \eqref{eq:sigmaA}, the most relevant eigenvalues are $-13\epsilon/8$ and $-\epsilon/2$ respectively. Combining these results with those for the $g_i$ in \eqref{eq:KTA}	we finally obtain,
	in place of \eqref{eq:sigmarescaling}, the following scalings, where for each term we have kept only the most relevant part:
	\begin{equation}\label{eq:finalsig}
	\begin{aligned}
	\sigma_1 &\to b^{d+z+5\epsilon/36}\sigma_1,\\
	\sigma_2 &\to b^{d+z}\sigma_2,\\
	\sigma_3 &\to b^{d+z-\epsilon/12}\sigma_3.
	\end{aligned}
	\end{equation}
	
	Of these, $\sigma_3$ is the leading term in the entropy production near criticality. It behaves as $\sigma_2\sim\xi^{-(d+z-\epsilon/12)}$, so that $\psi\sim \sigma \xi^{(d+z)}\sim \xi^{\epsilon/12}\sim \mt^{-\nu\epsilon/12}$. 
	We conclude that, generically, for AMA in $4-\epsilon$ dimensions, $\theta_\sigma= \nu\epsilon/12 + \mathcal{O}(\epsilon^2)$ is strictly positive. The critical entropy production is accordingly {\em larger} in magnitude than the natural scaling, by a factor that diverges as $\mt\to 0$. 
	
	The physics of a divergent reduced entropy production, $\psi\sim\mt^{-\theta_\sigma}$ with $\theta_\sigma > 0$, is not yet clear and merits further study. Anomalous dimensions $\theta$ such as $\theta_\sigma$ for physical quantities can in some cases be associated with a fractal dimension, such that the `mass' of that quantity within a spatial correlation volume scales as $m\sim\xi^\theta$. 
	This would imply that entropy is produced across a cascade of spatial scales between the correlation length $\xi$ and the short cutoff $\xi_0$. (The latter is always implicated when $\theta_\sigma \neq 0$ in the sense that $\psi\sim\mt^{-\theta_\sigma}\sim(\xi/\xi_0)^{\theta_\sigma/\nu}$.) While plausible, such a geometric interpretation is far from guaranteed, especially as $\theta_\sigma$ could equally signify anomalous scaling in the temporal rather than spatial domain.
	
	So far, we have not discussed the {\em sign} of $\sigma_3$. The operator average $\langle(\nabla\phi)^2\phi\nabla^2\phi\rangle_{\text{cr}}$ is of definite but unknown sign, while the sign of the
	prefactor in (\ref{eq:sigmaA}) depends on input parameters, because \eqref{eq:KTA} implies that neither $\lambda$ nor $\lambda+\kappa_1$ changes sign under the flow.
	A negative $\sigma$  is not excluded physically, since the RG calculation extracts only the critical part of the EPR and ignores large, $\mt$-independent positive contributions from the short-scale cutoff. (These can cause the total EPR to scale as $\sigma +\text{const.}\, \xi_0^{-(d+z)}$, say.) Nonetheless, the negativity of  $\sigma$ in some parameter regions requires further study and interpretation, perhaps along lines used elsewhere to explain sign-reversals of $\sigma$ on integrating out microscopic degrees of freedom such as chemical reactions \cite{Tomer}.
	
	In conclusion, further studies are needed to clarify the character and interpretation of the steady-state EPR, $\sigma$, close to the critical points of active field theories. Our work motivates such studies by showing that $\psi\sim \sigma \xi^{(d+z)}$ is governed by a nontrivial scaling exponent and can be finite, or even divergent, on approach to the critical point. This holds although the active models we studied are members of the conserved (Model B) and non-conserved (Model A) Ising universality classes, differing from other, passive members only by RG-irrelevant terms. An {\em exponent} for EPR should thus be viewed as part of the universal behavior of such classes, albeit with an {\em amplitude} that vanishes for all their reversible members. This `stealth entropy production' scenario is very different from one in which $\sigma$ vanishes fast enough that active class members become effectively passive at criticality. It would be interesting to seek examples of both scenarios in a wider range of models than those considered here. It would also be interesting to pursue numerical studies of the EPR, in the fashion of those done in \cite{Nardini2017} but close to criticality, which would shed some light on the nature of the fractal or other scaling properties of the EPR close to equilibrium-like fixed points.
	
	{\em Acknowledgments:}
	We thank Cesare Nardini for many useful discussions. FC is funded by EPSRC DTP IDS studentship, project number 1781654. Work funded in part by the European Research Council under the EU's Horizon 2020 Programme, grant number 760769. MEC is funded by the Royal Society.

	\pagebreak
	\widetext
	\begin{center}
		\textbf{\large Supplemental Material: \\ Stealth entropy production in active field theories near Ising-class critical points}
	\end{center}
\setcounter{equation}{0}
\setcounter{figure}{0}
\setcounter{table}{0}
\setcounter{page}{1}
\makeatletter
\renewcommand{\theequation}{\arabic{equation}}
\renewcommand{\thefigure}{\arabic{figure}}
\renewcommand{\bibnumfmt}[1]{[S#1]}
\renewcommand{\citenumfont}[1]{S#1}
The supplemental material consists of two parts and contains technical aspects of the calculations described in the main text. The first part is the calculation of the RG flow of the parameters $\lambda$ and $\kappa_1$, to get equation (11) of the main text. The second part is the calculation of the anomalous dimension of each operator that forms part of the entropy production.
Both parts are standard RG calculations, following the methods of \cite{Tauber2014sm} and \cite{AmitBooksm} for the first and second sections respectively.

\section{Computation of the flow of $\lambda,\kappa_1$ in Active Model A}\label{app:LN}

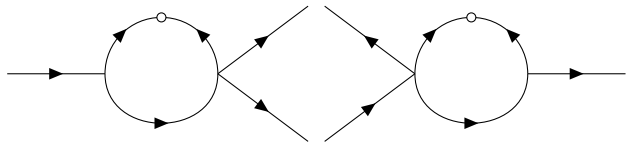
\begin{figure}[b]
	\centering
	\begin{tikzpicture}[baseline=-\the\dimexpr\fontdimen22\textfont2\relax]
	\begin{feynman}[inline=(a)]
	\vertex (a);
	\vertex[right=1.3cm of a] (b);
	\vertex[right=1.5cm of b] (c);
	\vertex[right=1.5cm of c] (d);
	\vertex[right=0.75cm of b] (e);
	\vertex[above=0.75cm of e] (f);
	\vertex[above=0.9cm of a] (g);
	\vertex[right=0.3 of g] (h);
	\vertex[below=0.9cm of a] (i);
	\vertex[right=0.3 of i] (j);
	
	\vertex[above=0.9cm of d] (g2);
	\vertex[left=0.3 of g2] (h2);
	\vertex[below=0.9cm of d] (i2);
	\vertex[left=0.3 of i2] (j2);
	
	\diagram* {
		(a) --[fermion] (b),
		(b) --[fermion, quarter left] (f),
		(c) --[fermion,quarter right] (f),
		(b) --[fermion, half right] (c),
		(c) --[fermion] (h2),
		(c) --[fermion] (j2),
	};
	\end{feynman}
	\draw[fill=white] (f) circle(0.6mm);
	\end{tikzpicture}
	\hspace*{0.1cm}\begin{tikzpicture}[baseline=-\the\dimexpr\fontdimen22\textfont2\relax]
	\begin{feynman}[inline=(a)]
	\vertex (a);
	\vertex[right=1.5cm of a] (b);
	\vertex[right=1.5cm of b] (c);
	\vertex[right=1.3cm of c] (d);
	\vertex[right=0.75cm of b] (e);
	\vertex[above=0.75cm of e] (f);
	\vertex[above=0.9cm of a] (g);
	\vertex[right=0.3 of g] (h);
	\vertex[below=0.9cm of a] (i);
	\vertex[right=0.3 of i] (j);
	
	\vertex[above=0.9cm of d] (g2);
	\vertex[left=0.3 of g2] (h2);
	\vertex[below=0.9cm of d] (i2);
	\vertex[left=0.3 of i2] (j2);
	
	\diagram* {
		(b) --[fermion, quarter left] (f),
		(c) --[fermion,quarter right] (f),
		(b) --[fermion, half right] (c),
		(b) --[fermion] (h),
		(j) --[fermion] (b),
		(c) --[fermion] (d),
	};
	\end{feynman}
	\draw[fill=white] (f) circle(0.6mm);
	\end{tikzpicture}
	\caption{\label{fig:lambdanu}One loop diagrams that contribute to $\lambda$ and $\kappa_1$ to order $\epsilon$. Lines are zeroth-order fields and the circle a bare correlator. The three-point vertices carry interactions $g_i$ and the four-point carries $u$.}
\end{figure}

This is the calculation of the RG flow of parameters of a dynamic theory. To order $\epsilon$, and above $2$ dimensions, so that both parameters remain irrelevant, we only need to calculate, to $1$ loop, the diagrams of Figure \ref{fig:lambdanu}. The computation is completely equivalent to the one done for the same diagrams in \cite{caballero2018sm}, except that in this computation the correlator and propagator are the ones of Active Model A (instead of Active Model B), for which we write the equation of motion according to the main text:
\begin{equation}
\dot\phi=-a\phi-u\phi^3+\kappa\nabla^2\phi-\lambda(\nabla\phi)^2+2\kappa_1\phi\nabla^2\phi+\eta.
\end{equation}

The loop integral will contain the propagator $G_0$ and correlator $C_0$ of the linear equation of motion, since we perturb in vanishingly small nonlinearities. These are
\begin{equation}
\begin{aligned}
G_0(q,\omega)&=\frac{1}{-i\omega+a+\kappa q^2},\\
C_0(q,\omega)&=\frac{2D}{\omega^2+(a+\kappa q^2)^2}.
\end{aligned}
\end{equation}

We use the following short-hand notation for the integral sign, which integrates out completely the time domain, and integrates the space domain in a thin shell close to the cutoff $\Lambda$:
\begin{equation}
\int_{k,\Omega}=\int_{-\infty}^{\infty}\frac{d\Omega}{2\pi}\int_{\Lambda/(1+db)}^\Lambda\frac{d^dk}{(2\pi)^d}.
\end{equation}
The loop integral for the sum of both diagrams can then be written as
\begin{equation}
-u\int_{k,\Omega}C_0(k,\Omega) \Big(g(q,k,q-k)G_0(q-k,-\Omega)+\frac{1}{2}g(q'-k,-k,q')G_0(q'-k,-\Omega)+\frac{1}{2}g(q-q'-k,-k,q-q')G_0(q-q'-k,-\Omega)\Big)
\end{equation}
where $g(k_1,k_2,k_3)$ is the vertex function
\begin{equation}
g(k_1,k_2,k_3) = -\kappa_1 (k_2^2+k_3^2)+\lambda k_2\cdot k_3,
\end{equation}
that comes from $\lambda$ and $\kappa_1$ terms in the Fourier transform of the equation of motion. The first term of the parenthesis comes from the first diagram, while the remaining two terms in $g(k_1,k_2,k_3)$ come from the second. (Do not confuse the notation $g$ for vertex functions here with the $g_i$ of the main text which are the coupling constants.)

The result of this integral, calculated for the interval of momenta we integrate out $(\Lambda/(1+db),\Lambda)$, and evaluated at zero mass, is
\begin{equation}
\frac{9 \bar u \kappa_1\Omega_d}{4 \Lambda^4} \left(q^2+2q'^2-2q\cdot q'\right)db+\frac{3 \bar u \lambda\Omega_d}{4\Lambda^4} \left(-q^2+2q'^2-2q\cdot q'\right)db,
\end{equation}
where $\Omega_d=S_d/(2\pi)^d$, with $S_d$ the surface of a $d$ dimensional sphere, and where $\bar u=Du\kappa^{-2}$ is the reduced interaction parameter in terms of which we write the flow. The fixed point is found for this parameter as $\bar u^*=\epsilon\Lambda^4/(9\Omega_d)$. The above expression can be rewritten, by manipulating the wavevector terms, as a sum of two terms in the form of the two terms of $g(k_1,k_2,k_3)$, so that we can identify the contributions to $\lambda$ and $\kappa_1$:
\begin{equation}
\frac{3 \bar u (3\kappa_1-\lambda)\Omega_d}{4 \Lambda^4} \left(q'^2+(q-q')^2\right)db-\frac{3 \bar u \lambda\Omega_d}{\Lambda^4} \left(q'\cdot(q-q')\right)db.
\end{equation}

Integrating over the wavenumber shell gives intermediate values for $\lambda$ and $\kappa_1$. These can be read off the above equation, whose first term contains the Fourier transform of $(\phi\nabla^2\phi)$ and the second that of $(\nabla\phi)^2$. Expanding also to $\epsilon$ order (which is a trivial expansion since both terms are proportional to $u$, whose fixed-point value is $\bar u^*=\Lambda^4\epsilon/(9\Omega_d)$) we obtain
\begin{equation}
\begin{aligned}
\lambda_I &= \lambda -\frac{\lambda\epsilon}{3}db,\\
\kappa_{1,I} &= \kappa_1 -\frac{\kappa_1\epsilon}{4}db+\frac{\lambda\epsilon}{12}db.
\end{aligned}
\end{equation}
Notice that, as usual, the flow does not depend on the cutoff $\Lambda$ when we work around the Wilson-Fisher fixed point, in the sense that the value of the fixed point itself depends on the cutoff, but not the scaling of operators.

From these intermediate values, the last step is to rescale the equation of motion using the appropriate scaling exponents, as described in the main text, to find the transformation of the parameters. After sending the scale factor $b\rightarrow 1+db+O(db^2)$ to obtain differential equations for the flow, we find
\begin{equation}
\begin{aligned}
\frac{d\lambda}{db} &= (z+\chi-2)\lambda-\frac{\lambda\epsilon}{3},\\
\frac{d\kappa_1}{db} &= (z+\chi-2)\kappa_1 -\frac{\kappa_1\epsilon}{4}+\frac{\lambda\epsilon}{12}.
\end{aligned}
\end{equation}
where the terms of order $\epsilon^0$ arise as the differential version of the scaling of the couplings $g_i$ in (9) of the main text, and the terms in $\epsilon^1$ follow directly from the intermediate values given above.

If we now rewrite this flow in terms of $\lambda$ and $\lambda + \kappa_1$, we obtain
\begin{equation}
\begin{aligned}
\frac{d\lambda}{db} &= (z+\chi-2)\lambda-\frac{\lambda\epsilon}{3},\\
\frac{d(\lambda+\kappa_1)}{db} &= (z+\chi-2)(\lambda+\kappa_1) -\frac{\epsilon}{4}(\lambda+\kappa_1),
\end{aligned}
\end{equation}
so that there is no longer any mixing between parameters. This allows  us to write equation (11) of the main text and obtain the corrections to the scaling of the entropy production. Note that the non-mixing between $\lambda+\kappa_1$ and other couplings is a necessary feature since  this parameter vanishes in equilibrium; therefore if it starts at zero, it must remain there under the RG flow.

\section{Computation of the anomalous dimensions of the composite operators}\label{app:LNa}

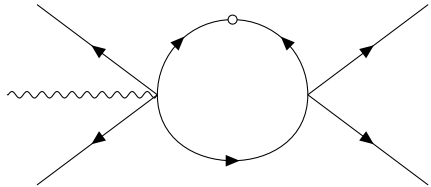
\begin{figure}
	\centering
	\begin{tikzpicture}[baseline=-\the\dimexpr\fontdimen22\textfont2\relax]
	\begin{feynman}[inline=(a)]
	\vertex (a);
	\vertex[right=2cm of a] (b);
	\vertex[right=2cm of b] (c);
	\vertex[right=2cm of c] (d);
	\vertex[right=1cm of b] (e);
	\vertex[above=1cm of e] (f);
	\vertex[above=1.2cm of a] (g);
	\vertex[right=0.4 of g] (h);
	\vertex[below=1.2cm of a] (i);
	\vertex[right=0.4 of i] (j);
	
	\vertex[above=1.2cm of d] (g2);
	\vertex[left=0.4 of g2] (h2);
	\vertex[below=1.2cm of d] (i2);
	\vertex[left=0.4 of i2] (j2);
	
	\diagram* {
		(a) --[photon] (b),
		(b) --[fermion, quarter left] (f),
		(c) --[fermion,quarter right] (f),
		(b) --[fermion, half right] (c),
		(b) --[fermion] (h),
		(b) --[fermion] (j),
		(c) --[fermion] (h2),
		(c) --[fermion] (j2),
	};
	\end{feynman}
	\draw[fill=white] (f) circle(0.6mm);
	\end{tikzpicture}
	\caption{\label{fig:oneloopcomp}One-loop diagram contributing to the coefficients $a_i$ to one loop order (to order $\epsilon$). The wavy line indicates the insertion of a composite operator.}
\end{figure}

As mentioned in the main text, to calculate the anomalous dimension of the operators of the entropy production, we need in principle to take into account all other operators that, in $4$ dimensions, have the same dimension as those featuring in the entropy production $\sigma$. To one loop (order $\epsilon$), there is mixing only among operators that separately contain the same number of gradients and fields, namely $4$ gradients and $4$ fields in the case of Active Model A. This comes from the fact that operators with more fields cannot, to one loop only, mix with a $u$ vertex to produce operators with less fields, and that operators with less than $4$ fields need more than one $u$ vertex to mix with in order to produce operators with $4$ fields, so when substituting $u$ with its value at the WF fixed point, we obtain terms of order higher than $\epsilon$. For an equivalent calculation done with operators of dimension $6$ in $4$ dimensions, see \cite{AmitBooksm}. This means that, to one loop, there is only one diagram to calculate, the one shown in Figure \ref{fig:oneloopcomp}, although the wavy line vertex represents several operators.

The number of linearly independent operators is seven. Mixing among these operators arises because, starting with the two that appear in the entropy production, each RG step generates all seven of them, so we need to calculate their flow equations as a coupled system. 

The operators in question are the following, written in real and Fourier space
\begin{equation}\label{eq:opdefs}
\begin{aligned}
A_1 &= \phi^3\nabla^4\phi\rightarrow\frac{1}{4}\sum q_i^4,\\
A_2 &= \phi^2\nabla\phi\cdot\nabla(\nabla^2\phi)\rightarrow \frac{1}{12}\sum q_i^2q_i\cdot q_j,\\
A_3 &=\phi^2\nabla^2\phi\nabla^2\phi\rightarrow \frac{1}{6}\sum q_i^2q_j^2,\\
A_4 &=\phi^2(\nabla_\alpha\nabla_\beta\phi)^2\rightarrow \frac{1}{6}\sum (q_i\cdot q_j)^2,\\
A_5 &=\phi\nabla^2\phi(\nabla\phi)^2\rightarrow \frac{1}{12}\sum q_i^2q_j\cdot q_k,\\
A_6 &=\phi\nabla_\alpha\nabla_\beta\phi\nabla_\alpha\phi\nabla_\beta\phi\rightarrow \frac{1}{12}\sum q_i\cdot q_j q_i\cdot q_k,\\
A_7 &=(\nabla\phi)^4 \rightarrow \frac{1}{3}\sum q_i\cdot q_j q_k\cdot q_l,
\end{aligned}
\end{equation}
where the sums are to be taken over the wavevector indices subject to a restriction that different indices must never take the same value. Note that the entropy production directly involves only $A_5$ and $A_7$.

The value of the diagram of Figure 2 of the main text contains the contribution of each of the seven operators to every other operator. If these are added to the action of the system with fields $a_i$: $\mathcal A\rightarrow \mathcal A+\int a_iA_i$, the one loop diagram defines a matrix $M_{ij}$, such that the intermediate values (after integration over eliminated momenta) can now be written
\begin{equation}
a_{i,I} = a_i+M_{ij}a_jdb.
\end{equation}
Here the matrix $M_{ij}$ is proportional to $u$, and so proportional to $\epsilon$ once it is evaluated at the fixed point, at which $\bar u = \bar u^*$ as discussed above.
The matrix $M_{ij}$ is as follows:
\begin{equation}
M_{ij}=\frac{\epsilon}{9}\left(
\begin{array}{ccccccc}
-9 & \frac{3}{2} & 0 & -3 & -\frac{3}{4} & \frac{3}{8} & 0 \\
0 & -\frac{9}{2} & 0 & -9 & 0 & \frac{3}{2} & -6 \\
0 & 0 & -3 & 0 & -\frac{3}{4} & -\frac{1}{8} & -1 \\
0 & 0 & 0 & -12 & 0 & 2 & -5 \\
0 & 0 & 0 & 1 & -\frac{9}{2} & -\frac{3}{2} & -4 \\
0 & 0 & 0 & -4 & 0 & -\frac{15}{2} & -2 \\
0 & 0 & 0 & 0 & 0 & -\frac{3}{4} & -6 \\
\end{array}
\right).
\end{equation}
we calculate explicitly one of the columns of $M_{ij}$ below. The rest of the calculation of $M_{ij}$ is similarly standard, but long, and will appear in \cite{thesissm}. As explained in the main text, the left eigenvectors of this matrix define the operators for which an anomalous dimension is well defined, as they are the directions in the space of these operators in which no other operator is generated. The decomposition of any arbitrary operator into these eigenvectors allows us to find the least irrelevant (in the RG sense) anomalous dimension involved in the scaling of any given operator \cite{AmitBooksm}.

Put differently, the RG flow for the composite operators must be analyzed by decomposing each $a_i$ as a sum of the eigenvectors of $M_{ij}$, each of which has a pure scaling. The leading-order scaling of each of the $a_i$, on approach to the fixed point, will be given by the least irrelevant among the eigenvectors for which the given $a_i$ has nonzero projection.

To illustrate the calculation of the matrix $M_{ij}$ we follows the standard treatment of \cite{AmitBooksm}, except that here instead of the free energy we use the Martin-Siggia-Rose action of the model, which is calculated from the Onsager-Machlup action (equation (2) of the main text) in also a completely standard way \cite{Tauber2014sm}. Therefore, the diagram in Figure 2 of the main text has one loop with one propagator and one correlator between Gaussian fields. The vertex itself is made of the Fourier transform of each of the seven operators $A_{1-7}$ introduced above. The idea is to calculate the diagram for each of the seven operators. Each will have a different low frequency expansion that can, in principle, be absorbed into either the operator itself or into any of the other six in some combination. A generic term in $M_{ij}$ gives the contribution of $A_j$ to $a_i$, so that each operator gives one column of the matrix. We show the computation here for $A_7$ as an example.

Since the vertex only mixes frequencies of the four outcoming fields $\phi$, we can write the diagram as follows,
\begin{equation}
\begin{aligned}
\frac{-36a_7u}{12}\int_{k,\Omega}&g(k,q_i,q_j)C_0(k,\Omega)G_0(-k-q_i-q_j,-\Omega)+\\
&\text{symmetrized terms},
\end{aligned}
\end{equation}
where the vertex function is now the one for $A_7$, also written in a symmetrized way
\begin{equation}
\begin{aligned}
g(k,q_1,q_2)=\frac{-1}{3}\Bigg(&k\cdot q_1 q_2\cdot(-k-q_1-q_2)+\\
&k\cdot q_2 q_1\cdot(-k-q_1-q_2)+\\
&q_1\cdot q_2 k\cdot(-k-q_1-q_2)\Bigg).
\end{aligned}
\end{equation}

Above, ``symmetrized terms'' refers to all terms that are produced by permuting the outcoming wavevectors $q_i$, and by doing the integral over the $\delta(q-\sum_iq_i)$ to write the correlator with either (i) the two wavevectors of the vertex of the insertion of $A_i$: $G_0(-k-q_i-q_j)$; or (ii) with the two wavevectors of the $u$ vertex: $G_0(-k+q_i+q_j)$. There are twelve terms after doing this, thus the $1/12$ at the front of the loop integral. Lastly, the $36$ is just the symmetry factor. This makes it easier to identify each resulting term with each of the $A_i$ operators.

After integrating the loop frequency, we obtain a series of terms, all proportional to $\bar ua_7\Omega_d\Lambda^{-4}db$, in which the four wavevectors $q_i$ are multiplied with each other in different combinations of dot products. Now we identify each term with each particular Fourier form of the $A_i$, for example, a term of the form $q_1^2q_2^2$ would be absorbed by $A_3$, and so would be part of $M_{3,7}$. We find this particular loop diagram to contribute
\begin{equation}
\bar ua_7\Omega_d\Lambda^{-4}db\Big(-6\tilde A_2-\tilde A_3-5\tilde A_4-4\tilde A_5-2\tilde A_6-6\tilde A_7\Big),
\end{equation}
where the tilde refers to the Fourier transform of each operator (see \eqref{eq:opdefs}).
After substituting $\bar u\rightarrow \bar u^* = \epsilon\Lambda^4/(9\Omega_d)$, we find the prefactor of each $\tilde A_i$ to be each term of the seventh column of $M_{ij}$. This has to be repeated for each operator to get each column of $M_{ij}$; see  \cite{thesissm} for details.

We finally return to the entropy production $\sigma$, where the decomposition of both operators of interest into the eigenvectors of $M_{ij}$  is just a linear algebra problem. Call $v^\alpha$ the eigenvector of each eigenvalue $\delta^\alpha$, where $\alpha$ takes values from $1$ to $7$. We label these in order of increasing (that is, decreasingly negative) eigenvalue; with this convention, 
the eigenvalues are readily found to be
\begin{equation}\label{eq:eigenvalues}
\delta^\alpha=-\epsilon\left(\frac{10}{9},1,1,\frac{13}{18},\frac{1}{2},\frac{1}{2},\frac{1}{3}\right).
\end{equation}

Also the  eigenvectors (not normalized) corresponding to each $\delta^\alpha$ are found as
\begin{equation}
\begin{aligned}
v^1 &= (0 , 0 , 0 , 2 , 0 , -1 , 2), \\
v^2 &= (-54 , 18 , 0 , -3 , -9 , 0 , 19), \\
v^3 &= (-84 , 28 , 0 , -30 , -14 , 19 , 0), \\
v^4 &= (0 , 0 , 0 , 8 , 0 , -11 , 36), \\
v^5 &= (0 , -11 , 0 , 12 , -9 , 0 , 28), \\
v^6 &= (0 , -33 , 0 , 36 , 1 , 7 , 0), \\
v^7 &= (0 , 0 , 28 , -2 , -14 , 1 , 12). \\
\end{aligned}
\end{equation}
Thus, to calculate the scalings of the composite operators, we need to know the eigenvector(s) onto which any given operators have nonzero projection. However their particular norm is unimportant, as are the numerical coefficients of the projection.

We arrive by this route at the following decomposition for the operators $A_5$ and $A_7$ that arise in the expression for the entropy production (Eq.~(8) of the main text):
\begin{equation}
\begin{aligned}
\tilde A_7 &= \frac{3}{7}v^1+\frac{-7}{95}v^2+\frac{9}{190}v^3+\frac{3}{70}v^4,\\
\tilde A_5 &= \frac{5}{7}v^1+\frac{-14}{95}v^2+\frac{9}{95}v^3+\frac{17}{140}v^4+\frac{-3}{28}v^5+\frac{1}{28}v^6.
\end{aligned}
\end{equation}
Having labelled the eigenvalues in order of decreasing negative magnitude, it is now the last nonzero term of each such decomposition that sets the leading (order $\epsilon$) correction to the scaling of each operator. The results in the main text for the scalings of $\sigma_{1,2,3}$ then follow directly.

We finish with a comment about the conserved model variants, Active Model B (AMB) and AMB+. As summarized in the main text, their calculation near the Gaussian fixed point is no harder than for AMA. The same is not true for the $\epsilon$-expansion of the entropy production rate which we do not pursue here. This is because, in AMB, $\sigma$ has dimensions of $\nabla^6\phi^4$, for which the operator mixing involves $20$ different operators instead of $7$. For AMB+ the same holds, except there is an additional active term entering the current equation that makes the entropy production non-local unless written as in terms of currents and not fields \cite{Nardini2017sm,tjhung2018reversesm}, making the $\epsilon$-expansion even more complicated. Since the ``stealth EPR" phenomenology of interest is already present in AMA for which the calculations, as outlined above, are more manageable, we have restricted ourselves to this model in the present work.

\end{document}